\documentclass[letterpaper,english,reprint,aps,fleqn]{revtex4-1}
\usepackage[latin9]{inputenc}
\usepackage{wrapfig}
\usepackage{units}
\usepackage{textcomp}
\usepackage{amsmath}
\usepackage{amssymb}
\usepackage{graphicx}
\usepackage{subscript}

\makeatletter

\pdfpageheight\paperheight
\pdfpagewidth\paperwidth

\DeclareFontEncoding{LGR}{}{}
\DeclareRobustCommand{\greektext}{%
  \fontencoding{LGR}\selectfont\def\encodingdefault{LGR}}
\DeclareRobustCommand{\textgreek}[1]{\leavevmode{\greektext #1}}
\ProvideTextCommand{\~}{LGR}[1]{\char126#1}

\usepackage{babel}

\usepackage{babel}

\makeatother

\usepackage{babel}
\begin{document}

\preprint{APS/123-QED}

\title{A Theoretical Basis for Epicatalysis Using Maxwell's Equations in
Their Quaternion Form:}

\author{Steven J. Silverman}

\affiliation{Cal State San Marcos, Department of Physics, San Marcos, California,
92096, U.S.A.}

\author{Nils Paz }

\affiliation{Independent Researcher, San Diego, California}

\author{John Harmon }

\affiliation{Independent , San Diego, California}

\date{\today}
\begin{abstract}
Recent experiments have demonstrated an interesting reaction on a
gas-surface defined as epicatalysis. The non-equilibrium thermodynamic
potentials were well described in a series of experiments. However
the theoretical basis was not established, based on Maxwell's electromagnetic
formulas. Therefore, this type of model was needed to explain the
experimental results. This paper suggests a connection between Maxwell's
fundamental electro-magnetic equations written in their quaternion
form and the thermodynamic measurements performed by Sheehan. The
connection to the thermodynamic variability is explained using missing
terms that the vector calculus interpretation of Maxwell's equations
over simplifies and ignores. As such, we provide a fundamental basis
to explain the associated results of epicatalysis. 
\begin{description}
\item [{GCAPS~numbers}] 05.20.Dd, 05.70.Ln, 05.70.Np, 88.05.De.{\small \par}
\end{description}
\end{abstract}

\pacs{73.20.Mf, 71.45.GM, 71.10.-w }

\keywords{Epicatalysis}
\maketitle

\section{\label{sec:level1}Introduction:}

Experimental work published in 2014 by Sheehan \citep{D.P.Sheehan2014},
demonstrates the possibility of maintaining two elements at two distinct
temperatures in a stationary state\footnote{In fact, a Non-Equilibrium Steady State whose state is maintained
but with changes of species across various interfaces.} without added external energy. The term for this phenomena was coined
as epicatalysis. Recent theoretical work has established an interesting
connection between Maxwell's equations, written in their original
quaternion form and thermodynamics. This connection is used to establish
a theoretical basis on the measured thermodynamic, non-equilibrium,
properties of epicatalysis that involes dissociation of Hydrogen gas
using Tungsten and Rhenium metals. The phenomenon emerges out of standard
(non quantum mechanical) kinetic theory in a limit of very low gas
pressure and density and strong gas-surface interactions. Due to the
differing desorption rates for monatomic and diatomic molecules (monomer
A and dimer A\textsubscript{2} respectively) (see Figure 1), there
arises differing pressures and temperatures between the faces of the
experimental apparatus vanes as shown in Fig 1. This can be harnessed
to perform work, in an apparent conflict with principles of thermodynamics.
The experimental realization of this paradox involves the dissociation
of low-pressure hydrogen gas on high-temperature refractory metals
(tungsten and rhenium) under black-body cavity conditions. The results,
corroborated by other laboratory studies and supported by theory,
confirm the paradoxical temperature differences and point to an explanation
through discarded terms in Maxwell's Equations ( Quaternionic Form
) as they are traditionally used in vector representation. In an attempt
to refute a newly predicted black-body cavity phenomenon, Duncan\citep{Duncan2000}
proposed: stationary-state pressure gradients arising due to surface-specific
dissociation and recombination of dimeric gases. Duncan's thought
experiment consists of a sealed black-body cavity housing a diatomic
gas (A\textsubscript{2}) and a radiometer. The opposing vane faces
have different chemical activities with respect to the gas-surface
reaction (A\textsubscript{2}, 2A),and different desorption and adsorption
rates for the dimer and the monomer. The experiments reported by Sheehan
\citep{D.P.Sheehan2014}, are the first experimental realization of
Duncan's thought paradox. The measurements, despite being conducted
under thermodynamically open non-black-body cavity conditions, served
to establish benchmark dissociation rates against which the results
of Duncan's black-body cavity experiments could be gauged. In Sheehan's
experiments, tungsten and rhenium filaments are the opposing metals
in a thermocouple. They were incorporated in two kinds of high-temperature
experiments. In Part I, the gas-filament experiments, the dissociation
rates were measured for both hydrogen and helium on refractory metal
filaments (tungsten and rhenium). They were quantitatively compared
over a range of temperatures and pressures (300 T - 2,000 K; 0.01
P 10 Torr). In Part II, the Duncan Paradox experiments, tungsten and
rhenium coated thermocouples were incorporated in high-temperature,
thermodynamically closed black-body cavities. Details of the experimental
setup are described in detail in reference \citep{D.P.Sheehan2014}.
In summary, the gas-filament and Duncan paradox results suggest modifications
and new interpretation to traditional theoretical understanding. If
the standard theory of heterogeneous catalysis is valid, then the
gas-filament experiment should not have been able to shift their H
and H\textsubscript{2} concentrations so far from equilibrium as
was observed, and not have 2 distinctly different values, in which
case they should not have been able to display different hydrogen
dissociation power consumption. This implies that $\bigtriangleup$P\textsubscript{hd}
should have been zero over the entire temperature-pressure range investigated.
These results can be explained, however, within a recently proposed theoretical extension to catalysis:epicatalysis \citep{X.Qi2003}. The Duncan Paradox experiments are more problematic. Within the traditional understanding of the thermodynamics,
standing temperature differentials such as those reported should not
be possible. The Duncan Paradox experiment, however, maintained two
distinct temperatures that did not relax to a single temperature because
its dual surface-specific reaction rates. Apparently, the Duncan Paradox
system constitutes a stationary-state of non-equilibrium. In addition,
the temperature differences in Duncan Paradox experiments generated
Seebeck voltages that can drive currents. In this paper, we analyze
this effect using Maxwell's equations in their original quaternion
representation.

\includegraphics[width=\linewidth]{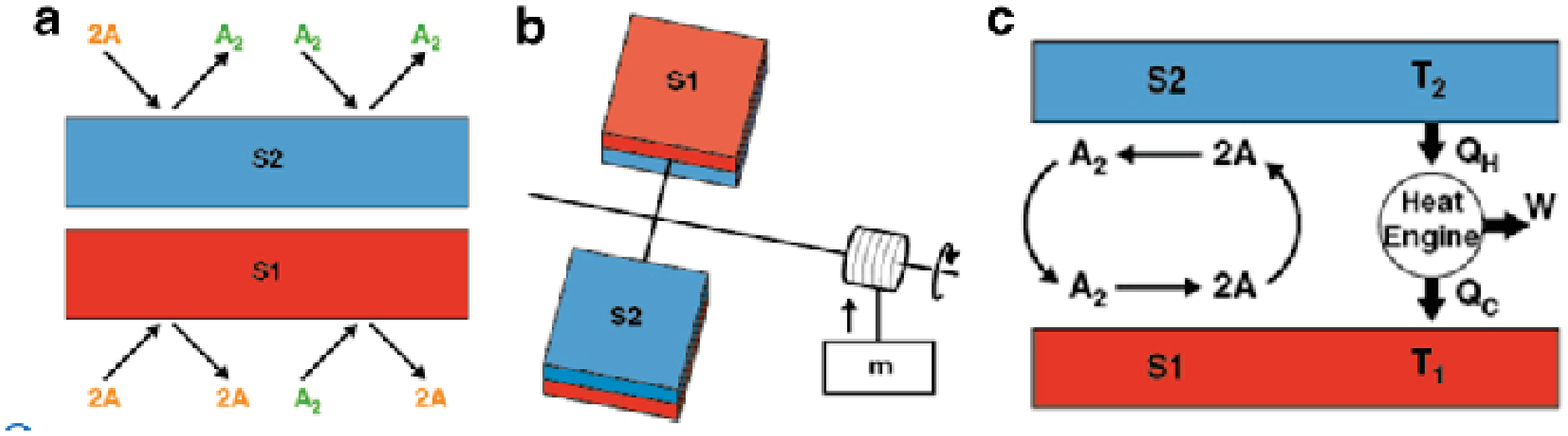}

Figure 1: Duncan's Paradox. 1a) A\textsubscript{2} is the diatomic
molecule. 2A is the monoatomic molecule. S2 is surface 2 and S1 is
surface 1. T\textsubscript{1} and T\textsubscript{2 }are the respective
temperatures for each surface. 1b) Shows a radiometer winding a mass
m. 1c) W is the work output for the cycle between the plates of the
2 surfaces\citep{D.P.Sheehan2014}.

\subsection{Experimental Graphs}

\begin{wrapfigure}{o}{0.5\columnwidth}%
\includegraphics[width=\linewidth]{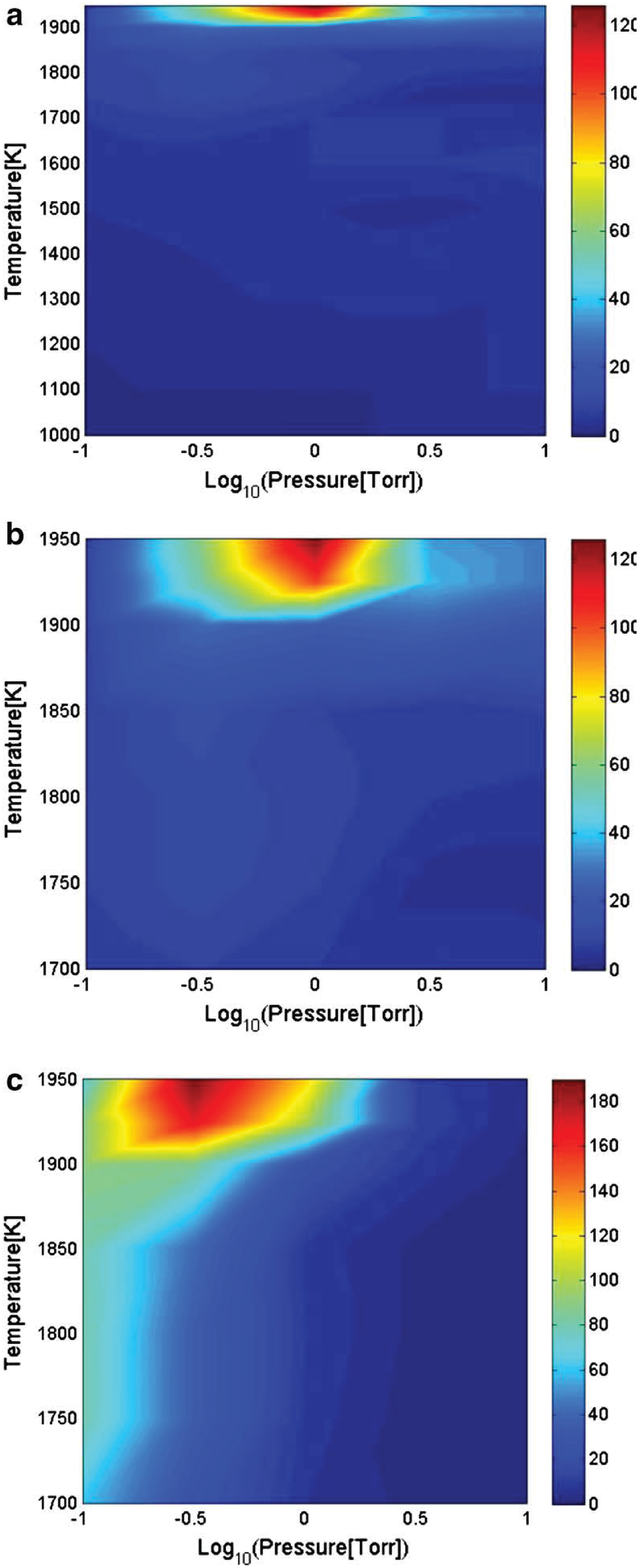}

\end{wrapfigure}%
Figure 2: Duncan Paradox experimental results. Duncan's temperature
difference. 2a) \textgreek{D}T\textsubscript{H$_{2}$\textminus Helium}
at experimental temperatures and pressures in tungsten black-body
cavity (1,000 \ensuremath{\le} T \ensuremath{\le} 1,950 K; 0.1 \ensuremath{\le}
P \ensuremath{\le} 10 Torr). Re cools relative to W. (Color scale
units: Kelvins (K).) Note: Additional measurements extend barren region
down to room temperature (290 K). 2b) (\textgreek{D}T\textsubscript{H$_{2}$\textminus He}/P)
in most active temperature interval of a (1,700 \ensuremath{\le}
T \ensuremath{\le} 1,950 K). (Color scale units: K.) 2c) Pressure-normalized
data from 2b (\textgreek{D}T\textsubscript{H$_{2}$\textminus He}/P)
versus experimental temperature and pressure. (Color scale units:
K/Torr) threshold \citep{D.P.Sheehan2014}

\section{Mathematical Developments}

In work by Jack\citep{Jack2003}Maxwell's equations are exhibited
in their original quaternion forms. Since this mathematical approach
may be unfamiliar to most physicists. We will summarize his results.
The interested reader is referred to his paper for more details.

A general quaternion is written as: 
\begin{eqnarray}
a=a_{0}+a_{1}i+a_{2}j+a_{3}k\label{eq:one}
\end{eqnarray}
where the i, j, k $($$i=\sqrt{-1},j=\sqrt{-1,}k=\sqrt{-1)}$ . where
i, j, k are the anti-commuting hyper complex roots of -1, and the
a0, a1, a2, a3 are elements of the real number set, can be used to
write down Maxwell's Equations. Then Maxwell electromagnetic potentials
(A) can be written as: 
\begin{equation}
A=U+A_{0}i+A_{1}j+A_{2j}
\end{equation}

The differential operator being written out in a similar manner as:
\begin{equation}
\frac{d}{dr}=\frac{d}{dt}+i\frac{d}{dx}+j\frac{d}{dy}+k\frac{d}{dz}
\end{equation}

With c = 1 (speed of light ). After some algebra we can state the
Electric and Magnetic field terms appear, as quaternions:

\begin{equation}
E=\left\{ \frac{d}{dr},A\right\} 
\end{equation}

\begin{equation}
B=\left[\frac{d}{dr},B\right]
\end{equation}

That is, the electric field is the negative symmetric derivative of
the potential, and the magnetic field is the positive anti-symmetric
derivative of the potential. 
\begin{equation}
E=-\nicefrac{1}{2}(\frac{d}{dr}\rightarrow A+A\leftarrow\frac{d}{dr})
\end{equation}

For the Electric Field (the arrows shown indicate multiplication right
or left). Then the Magnetic Field becomes:

\begin{equation}
B=+\nicefrac{1}{2}(\frac{d}{dr}\rightarrow A-A\leftarrow\frac{d}{dr})
\end{equation}

Which according to references \citep{W.R.Hamilton1843} the electric
field is the negative symmetric derivative of the potential, and the
magnetic field is the positive anti-symmetric derivative of the potential.
The space components of these quaternion fields correspond exactly
to the electric and magnetic fields in the usual 3-vector calculus.
However, the electric quaternion field now has a time component, which
we label, $\mathbf{\Omega}$, so that, E = \ensuremath{\mathbf{\Omega}}
+ E, while the magnetic quaternion field has no time component, so
that, B = 0+B. And if we allow our notation to alternate between Heaviside-Gibbs
3-vector and that of Hamilton's Quaternion 3-vector, taking care to
match up only the components of the appropriate expressions, we can
write the quaternion derivative in terms of the more familiar vector
notation as: 
\begin{equation}
\frac{d}{dr}\rightarrow A=\frac{\partial\phi}{\partial t}-\nabla\bullet A+\frac{\partial A}{\partial t}+\nabla\phi+\nabla\times A
\end{equation}
And for the other symmetry: 
\begin{equation}
\frac{d}{dr}\leftarrow A=\frac{\partial\phi}{\partial t}-\nabla\bullet A+\frac{\partial A}{\partial t}+\nabla\phi-\nabla\times A
\end{equation}

Then, by inspection, the reformulated Maxwell Field Equations are:
\begin{equation}
\left[\frac{d}{dr},B\right]=\left\{ \frac{d}{dr},A\right\} 
\end{equation}
and similarly 
\begin{equation}
\left[\frac{d}{dr},E\right]=-\left\{ \frac{d}{dr},B\right\} 
\end{equation}
The anti-symmetric derivative of the magnetic field is the positive
of the symmetric derivative of the electric field. And, the anti-symmetric
derivative of the electric field is the negative of the symmetric
derivative of the magnetic field. The first represents a real physical
law, while the second is easily proven to be an algebraic identity
when given the definitions of the electric and magnetic fields above.
The algebra dictates a new term. This term is not gauge transformed
away. In essence, it is an essential part of the equations whose existence
will lead us to a thermodynamic connection between Maxwell's equations
and electro-motive force (EMF). Writing the quaternionic equations
now in vector format with the more complete Maxwell Formulation for
macroscopic materials we have: 
\begin{equation}
\nabla\times H=\Bigl(\frac{1}{c}\Bigr)\Bigl(\frac{\partial D}{\partial t}+4\pi J\Bigr)+\bigtriangledown\text{\ensuremath{\mathbf{\Omega}}}
\end{equation}

\begin{equation}
\nabla\times D=-\frac{1}{c}\frac{\partial H}{\partial t}
\end{equation}

\begin{equation}
\nabla\bullet D=\left(\frac{1}{c}\right)\frac{\partial\mathbf{\Omega}}{\partial t}+4\pi\rho
\end{equation}

\begin{equation}
\mathbf{\nabla\bullet H=0}
\end{equation}

and the last set of equations we have: 
\begin{equation}
\mathbf{\Omega}=-\Bigl(\frac{1}{c}\Bigr)\frac{\partial\phi}{\partial t}+\nabla\bullet A
\end{equation}

\begin{equation}
D=-\bigtriangledown\phi-\Bigl(\frac{1}{c}\Bigr)\frac{\partial A}{\partial t}
\end{equation}

and finally: 
\begin{equation}
\mu H=\nabla\times A
\end{equation}
In this formalism, $D=\varepsilon E,$ $B=\mu H,$\citep{J.D.Jackson1975}and
provided we now identify the electric charge density $\rho$, and
electric current density, J, with the terms involving \ensuremath{\mathbf{\Omega}}.
Thus, $4\pi\rho$ =$\Bigl(\frac{1}{c}\Bigr)$ $\frac{\partial\mathbf{\Omega}}{\partial t}$,
and $\left(\frac{1}{c}\right)4\pi J$ = $\bigtriangledown$\ensuremath{\mathbf{\Omega}}.
According to\citep{Jack2003} the question of gauge and specifically
the Lorentz Gauge does not affect the temporal term \ensuremath{\mathbf{\Omega}}
in equations 12 and 16 above. Equation 17 is the familiar Lorentz
Gauge we arrive at: 
\begin{equation}
\nabla\bullet D=-\nabla^{2}\phi-\frac{1}{c}\frac{\partial}{\partial t}\nabla\bullet A=4\pi\rho+\frac{1}{c}\frac{\partial\text{\ensuremath{\mathbf{\Omega}}}}{\partial t}
\end{equation}

Then using equation 14 and 17 we have finally 
\[
-\nabla^{2}\phi-\frac{1}{c^{2}}\frac{\partial^{2}}{\partial t^{2}}\phi=4\pi\rho+\frac{1}{c}\frac{\partial\mathbf{\Omega}}{\partial t}
\]
\begin{equation}
-\nabla^{2}A-\frac{1}{c^{2}}\frac{\partial^{2}}{\partial t^{2}}A=\frac{4\pi J}{c}
\end{equation}
In addition we can derive the usual partial differential equation
for the vector potential A. The temporal term \ensuremath{\mathbf{\Omega}}
is not gauged away and is a basic part of the equation. Recall that
the scalar quantity \ensuremath{\mathbf{\Omega}} is the Temporal Field
and emerges from Maxwell's Equations in their Quaternionic form. The
scalar field $\phi$ has the same units-of-dimension as the electric
and magnetic vector fields in our Gaussian system of units. And thus,
for a given charge, q, the quantity, q\ensuremath{\mathbf{\Omega}},
has units of force, similar to the electric force, qE, and the magnetic
force, qv/c \text times {} B. However, this scalar force has no
space direction. Instead, it acts along the time line. These equations
then make a distinction between the ``thermal\textquotedblright{}
and the ``electric\textquotedblright{} source contributions to the
electromagnetic fields. According to P. W. Bridgeman\citep{P.W.Bridgeman1961}
that thermoelectric phenomena require the phenomenological description
of EMF to allow for two different kinds of electromotive forces, one
that provides what he calls the ``working\textquotedblright{} EMF,
and the other that provides the ``driving\textquotedblright{} EMF,
for the thermoelectric system. The ``working\textquotedblright{}
EMF is responsible for the production of the total energy that emerges
from the system, while the ``driving\textquotedblright{} EMF is responsible
for moving the charges in the system, giving rise to the electric
current. These two electro motive forces, traditionally considered
the same in electricity, are not the same when including thermoelectric
effects. In this case, over the given time interval, energy is absorbed
or evolved from the charge-field interacting system accordingly, as
the signs of the charge and the temporal field are the same or opposite.
Since this scalar energy does not require the charge to move in space,
in order to manifest as observable phenomena the energy that is absorbed
and/or evolved must be interpreted as a form of heat. Moreover, this
heat is proportional to the first power of the charge, and thus reverses
sign with the change in sign of the charge, or as a change in sign
of the electric current, making this a reversible heat engine, corresponding
to the experimental observations already known as Peltier and Thomson
Heats in thermoelectricity. It is this relationship that we will show
is directly responsible for temperature changes ( vis-a-vi e.m.f ),
that provide for explanation of experimental results showing extensions
to the current interpretation of thermodynamic principles. 

\section{Thermodynamic Interpretation of Maxwell's Fundamental Equations}

In this section we summarize the equations as developed by \citep{Jack2003}Jack
, refer to his paper for more details. It is the temporal field, \ensuremath{\mathbf{\Omega}}
, that represents the total heat energy per unit charge evolving per
unit time (i.e. time measured in units of length) from the charge-field
interacting system due to both the loss in electrostatic potential
energy at the location and the flow of electrodynamic momentum out
of the same location, much like the diffusion heat flow being due
to the sum of two different processes, conduction and convection,
in non-equilibrium thermodynamics. This relationship between the temporal
field and temperature can be more clearly expressed functionally,
\ensuremath{\mathbf{\Omega}} = \ensuremath{\mathbf{\Omega}}(T). And
with this in mind, we conclude that the time rate of change of the
temporal field is proportional to the time rate of change of the temperature:
We note the following thermodynamic relationships\footnote{Here we return to the form of the electric and magnetic field in a
microscopic sense }: 
\begin{equation}
\bigtriangledown\bullet E_{\mathbf{\Omega}}=+\frac{1}{c}\frac{\partial\mathbf{\Omega}}{\partial t}
\end{equation}
The time derivative of the temporal field in equation 21 plays the
role of a charge density. We can derive a different form for the divergence
equation here much like we do for the electromagnetic field in materials
where we use the symbol D to indicate the effective electric field
in media. In this format we preserve this relationship as: 
\begin{equation}
\nabla\bullet(E-E_{\mathbf{\Omega}})=4\pi\rho
\end{equation}

Given the thermodynamic relationship to heat capacity $Q=m$C$\bigtriangleup(Temperature$),
we conclude that the time rate of change of the temporal field is
proportional to the time rate of change of the temperature, through
the relationship, $\frac{\partial\mathbf{\Omega}}{\partial t}=\frac{\partial\mathbf{\Omega}}{\partial T}\frac{\partial T}{\partial t},$where
we adapt the notation of Jack, with T being the thermodynamic Temperature.
We can then make the connection to Heat Capacity (C) and the Temporal
Field as: 
\begin{equation}
\frac{\partial\mathbf{\Omega}}{\partial T}=-\frac{1}{qc}\frac{d(-qc\mathbf{\Omega})}{dT}=C_{q}
\end{equation}

The speed of light now restored, and with $-qc\text{\ensuremath{\mathbf{\Omega}}}$
the Heat energy absorbed per unit time by the charge q. As such it
plays the role of material dependant heat capacity. Then the above
equation for the change of Temporal field \ensuremath{\mathbf{\Omega}}
with respect to time t is further interpreted as: 
\begin{equation}
\frac{\partial\mathbf{\Omega}}{\partial t}=C\frac{\partial T}{\partial t}
\end{equation}

In essence we have integrating over time t: 
\begin{equation}
Q=\Delta\mathbf{\Omega}(T)=C_{q}\Delta T
\end{equation}

Thus, the temporal field, \ensuremath{\mathbf{\Omega}} , represents
the total heat energy per unit charge (Q) evolving per unit time (i.e.
time measured in units of length) from the charge-field interacting
system due to both the loss in electrostatic potential energy at the
location and the flow of electrodynamic momentum out of the same location,
much like the diffusion heat flow being due to the sum of two different
processes. Refer to equation (19)and we develop a Gauss Law type relationship
with $E_{\mathbf{\Omega}}:$ 
\begin{equation}
\int\int\int\bigtriangledown\bullet E_{\mathbf{\Omega}}dV=\frac{4\pi}{\varepsilon(\varpi)}\sigma_{ENCLOSED}
\end{equation}

Or for our two dimensional geometry: 
\begin{equation}
E_{\mathbf{\Omega}}=\frac{\sigma_{Enclosed}}{\epsilon(\omega)}
\end{equation}

a familiar result for a metal conductor. Now we relate this formula
to equation 19 and we get: 
\begin{equation}
\frac{\sigma_{enclosed}}{\epsilon(\varpi)}=C_{q}\frac{\partial\mathbf{\Omega}}{\partial t}
\end{equation}

Up to this point we have shown many equations of electromagnetsim
that show the functional changes that occur when we incorporate the
temporal filed . A final set of equations can be constructed that
will prove valuable to our discussion.

\section{Results and Discussion}

We see that this novel usage of quaternion form of Maxwell Equations
leads to a more comprehensive thermodynamic interpretation of epicatalysis.
We base our work on several papers. As we seek to relate the power
( disassociate ) difference to Rhenium and Tungsten as shown in the
results  \citep{D.P.Sheehan2001}and \citep{D.P.Sheehan2014},we will
keep the time derivatives and relate them to the quantities Sheehan
has measured. As such we will make the association as: 
\begin{equation}
P_{H/He}=\frac{\partial\mathbf{\Omega}(T)}{\partial t}=C_{H/He}\frac{\partial T}{\partial t}
\end{equation}

Where the subscript H or He refers to either types of atomic elements.
As a result of differing desorption rates for A and A\textsubscript{2}
, there arise permanent pressure and temperature differences, either
of which can be harnessed to perform work. These are unexpected results
obtained by Sheehan's experiments. Here we report on the first theoretical
support of this experimental realization, involving the dissociation
of low-pressure hydrogen gas on high-temperature refractory metals
(tungsten and rhenium) under black-body cavity conditions. The results,
corroborated by other laboratory studies and supported by this theory,
confirm the paradoxical temperature difference and point to physics
beyond the traditional understanding of the second law

\includegraphics[width=\linewidth]{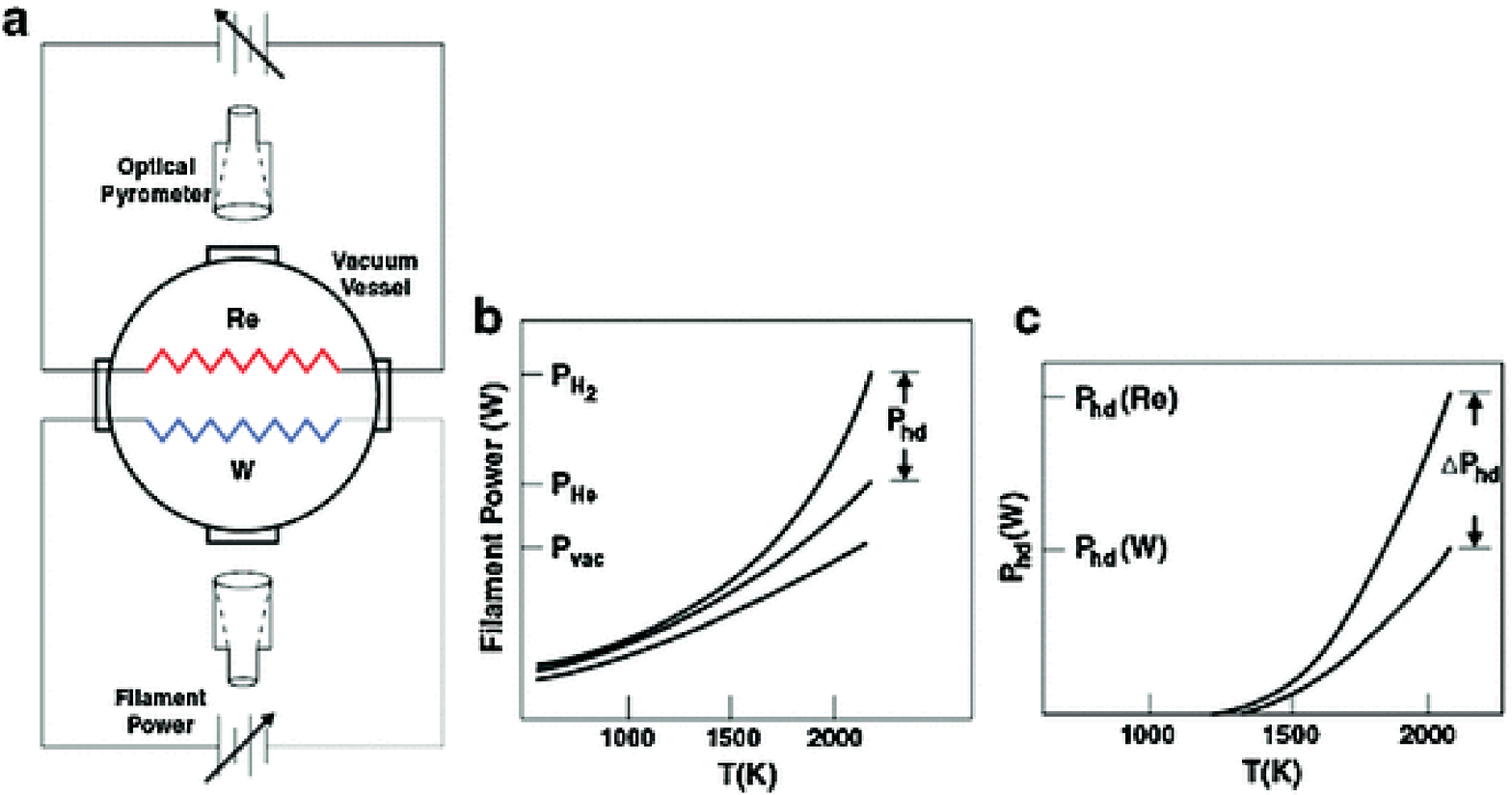}

Figure 3: GF experiment: 3a) A Hydrogen dissociation apparatus. Filaments
symbolized by resistors. 3b) Filament power versus temperature under
vacuum and under identical pressures of He and H2. Hydrogen dissociation
power: P$_{hd}$ =P$_{h2}$\textminus P$_{he}$. 3c) Hydrogen dissociation
power (P$_{hd}$) for Re and W filaments. \textgreek{D}P$_{hd}$ is
the difference between P$_{hd}$ (Re) \textminus P$_{hd}$(W) (H2
dissociation rates between these two metal filaments)\citep{D.P.Sheehan2014}. 

Several interpretations are possible to relate this power difference
to Maxwells Equations with the temporal field. The first relation
can be exhibited using $-qc\text{\ensuremath{\mathbf{\Omega}}}$ the
Heat energy absorbed per unit time by the charge q as indicated above.
In this guise we could write: 
\begin{equation}
-qc\triangle\mathbf{\Omega}=\triangle P_{hd}
\end{equation}

or 
\begin{equation}
-qc\left(\mathbf{\Omega}^{RE}-\mathbf{\Omega}^{W}\right)=P_{hd}(Re)\text{\textminus}P_{hd}(W)
\end{equation}

This form does not exhibit the material properties that are needed
on the left hand side (LHS) in order to be predictive. That is to
formulate the left had side of equation 31 in terms of Rhenium and
Tungsten. Instead it is more interesting to manipulate Maxwell Equations,
specifically equation 12. We do this by multiplying this equation
by the ratio of current to conductivity, $J$/\textgreek{sv}. This
will include the material specifications we require for this case
of Re and W but also for other metals substances that could exhibit
the same phenomena. This equation (31) can be rearranged and interpreted
as Heat Loss or Heat Gain by the metal differences, again see \citep{Jack2003}equation
37 in his work. We arrive at the suggestive form: 
\begin{equation}
\frac{dQ}{dt}=J^{2}/\sigma+\frac{c}{4\pi\sigma}\frac{d\mathbf{\Omega}}{dK}J\centerdot\nabla T
\end{equation}

$\frac{dQ}{dt}$ being the rate or loss of heat from the system is
the sum of two terms. One being the Joule Heat and the other being
the reversible Thompson like heat. This form fits in well with the
interpretation of $\triangle P_{hd}$. We propose a linear model relationship
between \ensuremath{\mathbf{\Omega}} T Thermodynamic Temperature.
Similiar in philosophy to the relationship between gravitational mass
and inertial mass from mechanics. In this sense we have an equation
derived from 32 for each metal. That is the difference: 
\begin{equation}
\frac{dQ}{dt}_{RE}-\frac{dQ}{dt}_{W}=J^{2}\left\{ \frac{1}{\sigma_{RE}}-\frac{1}{\sigma_{W}}\right\} +\frac{c}{4\pi}A\triangle TJ\centerdot\nabla T
\end{equation}

We expose here the difference in thermodynamic temperature between
the two metal plates from the Sheehan experiment. In the gradient
of \ensuremath{\mathbf{\Omega}} we assume it to be the same across
both plates. This is relative to the same geometry in both metals.
(change in temperature T over x and y directions should follow the
same relationship ). In essence geometry will not effect this term
since there are no differences between the metal plate geometry. However
the \textgreek{D}T ( change in thermodynamic temperature ) reflects
no geometry effect. The differences that might exist are subtracted
out due to geometry. In addition we assume a model \ensuremath{\mathbf{\Omega}}$_{}=A$(T).
The term A is a constant. We can fit that to the experiment in question
or just set A=1 for a first case. Combining equations 33 and 32, our
final result:
\begin{equation}
\begin{split} & J\left\{ \frac{1}{\sigma_{RE}}-\frac{1}{\sigma_{W}}\right\} +\frac{c}{4\pi}A\left\{ \frac{K_{RE}}{\sigma_{RE}}-\frac{K_{W}}{\sigma_{W}}\right\} J\centerdot\nabla T\\
 & =P_{hd}(Re)\text{\textminus}P_{hd}(W)
\end{split}
\end{equation}

What we need is the change in thermodynamic temperature on the LHS
to be related to measurable effects of the right hand side (RHS) in
34 above. The electrical conductivity of Rhenium and Tungsten is measured
as $\sigma_{RE}=5.6\text{\texttimes}10^{6}S/m$, while for Tungsten
we have $\sigma_{W}=2\text{\texttimes}10^{7}S/m$. The J$^{2}$ term
is always positive. However its product with the inverse conductivity
differences is less than zero. In the case where the gradient of T
= 0, (no gradient of temperature across surface ) .\footnote{Even if the gradient of T across the geometry is non zero, we can
easily show that its dot product with J which is essentially a surface
current close to 0.}which would represent non-equilibrium conditions and not steady state.
We find from Sheehan $P_{hd}(Re)\text{\textminus}P_{hd}(W)$= -0.8
W in agreement with our prediction. Other metals would show varying
differences as well.

This provides a clear modeling approach for this particular phenomena
and future experimental branching. We intend to explore this approach
to other experimental prototypes.
\begin{acknowledgments}
We wish to acknowledge the support of Professor Daniel Sheehan for
offering suggestions and encouragement in this work. 
\end{acknowledgments}

\appendix

\end{document}